\documentclass[twocolumn,preprintnumbers,amsmath,prc,amssymb]{revtex4}

\usepackage{graphicx}
\usepackage{dcolumn}
\usepackage{bm}
\usepackage{amsmath}
\usepackage{lipsum}

\begin{document}

\title{Self-consistent  Treatment of Quantum Gases of D-dimensional Hard Spheres Beyond the Van der Waals Approximation}

\author{K. A. Bugaev$^{1,2}$}

\affiliation{ $^1$Bogolyubov Institute for Theoretical Physics,
Metrologichna str. 14$^B$, Kiev 03680, Ukraine}
\affiliation{$^2$Department of Physics, Taras Shevchenko National University of Kiev, 03022 Kiev, Ukraine
}


\begin{abstract}
The necessary conditions to derive the quantum VdW EoS with hard-core repulsion from the quantum partition are discussed.  On a plausible example it is shown that an alternative way to account  correctly for the 3-rd virial coefficient  of classical hard spheres  leads to inconsistencies. The multicomponent formulation of the quantum VdW EoS with hard-core repulsion is derived within a self-consisting approximation. For practical applications   it is simplified, extended to higher densities and generalized to the case of   hard convex bodies of any dimension $D\ge 2$.\\

\end{abstract}

\maketitle

\section{Introduction} \label{Intro}

The concept of  hard-core repulsion plays an important role in the statistical mechanics of classical systems since 
it {  allows  one} to correctly account for the basic properties of real gases at short distances despite of its simplicity.
{Furthermore, the systems of hard spheres and hard discs have played the role of test sites to demonstrate the power of novel mathematical methods and new ideas in physics for decades.}
Since the invention of  the Van der Waals (VdW) equation of state (EoS) \cite{VDWeos} a great progress in extending the hard spheres  EoS to the high packing fractions {has been achieved} \cite{Ref2,Ref3}. 

Unfortunately, this is not the case for quantum systems, since the rigorous quantum statistical treatment of dense systems is extremely difficult.  
{Moreover, even  for a single sort of particles  the  quantum VdW EoS with the hard-core repulsion was never obtained from the quantum  grand canonical partition and the conditions that are necessary  for such a derivation were never clearly stated and discussed. 
However, the intensive  experimental studies of nuclear \cite{Sagun14,Aleksei18} and hadronic \cite{IST2018,IST2018b,Universe2019} systems 
and  other phases of strongly interacting matter \cite{3CEP}
require the quantum  EoS with multicomponent hard-core repulsion, i.e. with many different hard-core radii of particles.  
Besides, the  realistic quantum EoS are absolutely necessary to model the properties of dense quantum systems which can be nowadays studied  in atomic traps. 
}

Recently the situation is improving, since 
 presently {there exist  two quantum  EoS which, in principle,   allow one to go beyond the second  virial coefficient, but  none of them was,
 so far,  obtained
from the quantum partition function. The first of these  quantum EoS suggested in Ref. \cite{Vovch17} is a generalization of a famous 
one-component Carnahan-Starling EoS \cite{CSeos}. However, it was obtained in the canonical ensemble by replacing the total volume of the system $V$ by the volume $V_{av}$ that  is available for the motion of  $N$ particles  in a  free energy of some abstract  system.  
Despite its thermodynamic consistency, in the grand canonical ensemble  the  generalization of  Carnahan-Starling  EoS suggested in Ref. \cite{Vovch17} looks somewhat  unnatural, since it
contains not only the native variables of this ensemble, i.e. the temperature $T$ and chemical potential $\mu$, but also the particle number density. Moreover, below it will be shown that without additional assumptions it cannot be derived from the quantum partition.

Another quantum EoS of such type  is based on the concept of  induced surface tension  (IST) \cite{Aleksei18,KABugaev18} which allows one to account for the second, third and fourth virial coefficients of classical hard spheres \cite{IST2018}.   In contrast to the quantum generalization of  Carnahan-Starling  EoS of Ref.  \cite{Vovch17},  the IST one is formulated solely in the grand canonical variables and from the very beginning it is written for a mixture of an  arbitrary number of  sorts of particles  \cite{Sagun14,KABugaev18}. The IST EoS can be derived in two ways (compare the derivations of Refs.  \cite{Sagun14} and  \cite{Aleksei18}), but both of them are heuristic. In particular, in Ref. \cite{Aleksei18} the IST EoS was derived in the grand canonical ensemble using the generalization of a popular trick to replace the chemical potential $\mu$ by a shifted  one $\mu \longrightarrow \mu - b p$ \cite{Dirk91}, where $b$ is the excluded volume of a pair of particles and $p$ is the system pressure.

Unfortunately, the absence of  a  rigorous mathematical procedure on how to generalize the classical EoS  to the quantum system leads to  the appearance of  the statements that the usual VdW procedure by replacing the total volume of the system by the available one (and its equivalent in the grand canonical ensemble $\mu \longrightarrow \mu - b p$ suggested in  \cite{Dirk91} and generalized in \cite{Aleksei18,KABugaev18})  is not a unique one \cite{Typel16}. Althrough there was a strong critique of such claims \cite{KABugaev18}, it is mathematically involved and, therefore not widely available. Thus,  a rigorous derivation of the quantum VdW EoS  and its multicomponent generalizations like the quantum IST EoS \cite{Aleksei18,KABugaev18} are absolutely necessary, since these are  the basic tasks   of the quantum statistics 
of interacting particles. Moreover, 
 due to the fast  development of atomic traps and the corresponding possibilities to study the EoS of dense quantum systems such analysis is highly desirable. 

The work is organized as follows. In Sect.~2 the quantum VdW EoS is obtained by the Laplace transform me\-thod from quantum grand canonical partition. The pitfalls of the alternative ``derivation" to account for the third virial coefficient of classical hard spheres in a quantum EoS  are closely   discussed in Sect.~3. 
The 
detailed
 analysis of the excluded volume of the mixture of hard spheres of different radii and the corresponding  EoS of hard spheres and hard convex particles are given  in Sect.~4, while Sect.~5 is devoted to conclusions.

\section{2. Brief  Derivation of Quantum VdW EoS for Hard Spheres} \label{Model_vdw}

In all textbooks on statistical mechanics it is written that the VdW EoS for Boltzmann statistics cannot be rigorously derived, since it is 
an extrapolation of the low density expansion to high densities. Therefore,  it is of great interest and importance to derive the quantum VdW EoS 
for a gas with hard-core repulsion.
In contrast to the heuristic treatment of the quantum VdW EoS of Refs. \cite{Qvdw1,Qvdw2} here, for simplicity,  we consider only the hard-core repulsion and start  the discussion with the energy spectrum of the system of one sort of particles. The particles  with a hard-core repulsion, classical or quantum, behave as an ideal gas since the particles that do not touch each other have zero potential energy, whereas in the case 
when they touch each other their potential energy is infinite, but such states cannot contribute into the partition. Hence, we neglect the effects of Lorentz contraction of the eigen volume $V_0$ of  the particles \cite{RelVDW1,RelVDW2} and  consider the dilute systems and low temperatures compared to the particle mass $m$.

Suppose that $\{e_k\}$ is a set of all possible energies of a single particle. Then the total energy of  an  ideal gas with hard-core repulsion 
is $E= \sum_k^\prime n_k e_k$ and the total number of particles of the system of volume $V$ is $N= \sum_k^\prime n_k$, where the primed 
summations denote the sum not over the energy levels, but over the different states of a particle. Hence, if the  level with energy $e_k$ has a degeneracy $g_k$, then in the primed summations such a level, corresponding to the energy $e_k$, is taken into account exactly $g_k$ times. 
Now we can write the quantum partition of such a system as 
\begin{eqnarray}\label{Eq1}
&\hspace*{-3.3mm}Q (T, \mu, V) = \hspace*{-1.1mm}\sum\limits_{N=0}^\infty \sum\limits_{n_1+n_2+...=N} \hspace*{-1.1mm}e^{\left[ \frac{\mu N - E}{T} \right]} \Theta\left(V - bN \right)=\hspace*{-1.1mm} \sum\limits_{N=0}^\infty\nonumber \\
&\hspace*{-5.5mm}  \sum\limits_{n_1+n_2+...=N} \hspace*{-1.1mm}e^{\left[ \frac{\mu (n_1+n_2+...) - (n_1e_1+n_2e_2+...)}{T} \right]} \Theta\left(V - bN \right)\,. 
\end{eqnarray}
Here $\mu$ is the chemical potential of the system, while $T$ is its temperature and $b=4\,V_0$ denotes the excluded  volume per pair of particles (classical second virial coefficient), while $V_0$ is an eigen volume. The $\Theta$-function in Eq. (\ref{Eq1}) allows one to formally extend the summation over $N$ to infinity, but in fact it restricts such a sum to a finite value of $\max \{N \}= [V/b]$. This is the first simplification of the VdW approximation, since the dense packing of particles 
corresponds to a larger number $N_{dens} \simeq [0.741 V/V_0] \sim 3 \max \{N \}$, but  if we use the correct number $N_{dens}$ the low density limit will be wrong (see below). In order to find the correct thermodynamic limit of the partition (\ref{Eq1}) we use the technique of the Laplace transform with respect to $V$ to the isobaric partition \cite{Reuter08} 
\begin{eqnarray}\label{Eq2}
&\hspace*{-7.7mm}{\cal Q} (T, \mu, \lambda) = \int\limits_0^\infty d V e^{-\lambda V} Q (T, \mu, V)  = \int\limits_0^\infty d V e^{-\lambda V} \times \nonumber  \\
&\hspace*{-3.3mm} \sum\limits_{n_1} \sum\limits_{n_2} ...   e^{\frac{(\mu-e_1) n_1}{T}} e^{\frac{(\mu-e_2) n_2}{T}} ...
 \Theta\left(V - b(n_1+n_2+...) \right)= \nonumber \\
&\hspace*{-5.5mm}  \int\limits_0^\infty d \tilde V e^{\lambda \tilde V} \sum\limits_{n_1} \sum\limits_{n_2} ...   e^{\frac{(\mu-e_1-Tb\lambda) n_1}{T}} e^{\frac{(\mu-e_2-Tb\lambda) n_2}{T}} ...
 \Theta\left(\tilde V \right)\,, \,\,
\end{eqnarray}
where in the last equation above we changed the variable $V$ to the available volume $\tilde V = V - b(n_1+n_2+...)$ and accounted for the extra terms 
in the  exponential functions  with a corresponding value of $n_k$. This change of variable  allows us to sum  over the  values of $n_k$ to get rid of the  $\Theta$-function and to express  the partition as a product over all levels of energy 
\begin{eqnarray}\label{Eq3}
&&\hspace*{-2.5mm}{\cal Q} (T, \mu, \lambda) =  \nonumber \\
&& \int\limits_0^\infty d \tilde V \exp \left[   \sum\limits_k   \ln  \left[\sum\limits_{n_k}  e^{\frac{(\mu-e_k-Tb\lambda) n_k}{T}}\right] -\lambda \tilde V  \right]  
\,. \,~
\end{eqnarray}
 Henceforth, for definiteness' sake, we  will mainly analyse the systems with the Fermi-Dirac  statistics (FD), but  we will  also give the answers for the Bose-Einstein  (BE) one in the appropriate places.  Hence, the sum over $n_k$ under the logarithm symbol can be written as
\begin{eqnarray}\label{Eq4}
&&\hspace*{-2.5mm}
\sum\limits_{n_k=0}  e^{\frac{(\mu-e_k-Tb\lambda) n_k}{T}} \equiv \sum\limits_{n_k=0}  \omega(k, \lambda)^{n_k} =  \nonumber \\
&& \left\{
 \begin{array}{ll}
 1+ \omega (k, \lambda) \,, & ~{\rm for ~FD},\\
\left[1 - \omega (k, \lambda)\right]^{-1} \,, &  ~{\rm for ~BE}\,.
\end{array}
\right.
%
\end{eqnarray}
In order to make the integration over $\tilde V$ we have to make the usual Thomas--Fermi approximation for the density of discrete levels of energy (density of states) \cite{Huang} 
\begin{eqnarray}\label{Eq5}
&&\hspace*{-11mm}
 \sum\limits_k  \ln  \left[1 \pm \omega (k, \lambda) \right] \rightarrow \tilde V \hspace*{-0.1mm} \int \hspace*{-0.1mm}\frac{d^3 k  \, g_k}{(2 \pi \hbar)^3}   \ln  \left[1 \pm \omega (k, \lambda) \right] 
\,. \, 
\end{eqnarray}
Note, however, that compared  to the ideal gas case considered in textbooks 
the volume of the system in Eq. (\ref{Eq5}) has to be replaced by the available volume $\tilde V$. 
A word of caution  has to be said here with respect to the degeneracy: for low densities the prescription (\ref{Eq5}) with $g_k = g=  const$ is reasonable, but for high densities it is possible  that one has to assume an energy dependence of $g_k$.  Actually, one of a few principle purposes to make this analysis  was a necessity to motivate the experimental work  to study the hard-core repulsion in quantum systems  at high  densities 
in order to clarify the possible energy dependence of $g_k$.

Assuming in this work that $g_k = g= const$ is the spin-isospin degeneracy of considered particles  we can make integration in Eq. (\ref{Eq3}) and get the isobaric partition
\begin{eqnarray}\label{Eq6}
&& \hspace*{-2.5mm}{\cal Q} (T, \mu, \lambda) = 
\left[\lambda \mp g \int \frac{d^3 k  \, }{(2 \pi \hbar)^3}   \ln  \left[1 \pm \omega (k, \lambda) \right]   \right]  
\,, \,~
\end{eqnarray}
for the model one (M1, hereafter). In what follows the upper (lower) sign will be for FD (BE) statistics. Similarly to the analysis of Ref. \cite{Reuter08} it is easy to show that in the thermodynamic limit $V \rightarrow \infty$  the rightmost singularity of the partition (\ref{Eq6})  is the simple  pole $\lambda_0$ defined as 
 \begin{equation}\label{Eq7}
\lambda_0 = \pm g\int \frac{d^3 k }{(2 \pi \hbar)^3}   \ln  \left[1 \pm \omega (k, \lambda_0) \right] ,
\end{equation}
which is located   on the real axis in the complex plane of $\lambda$.
Then after  making the inverse Laplace transform 
\begin{eqnarray}\label{Eq8}
&& \hspace*{-9.5mm}Q (T, \mu, V) = 
\frac{1}{2 \pi i}\int\limits_{\chi -i\infty}^{\chi +i\infty} d \lambda\, {\cal Q} (T, \mu, \lambda ) \, \exp(\lambda V) \biggr|_{V\rightarrow\infty}  \hspace*{-5.5mm}\longrightarrow \nonumber \\
&& \hspace*{-9.5mm} \exp(\lambda_0 V) \left[1-  b\, g \int \frac{d^3 k  \, }{(2 \pi \hbar)^3}  \frac{1}{[\omega (k, \lambda_0)]^{-1} \pm 1} \right]^{-1}
\,, \,~
\end{eqnarray}
where the integration contour in the complex $\lambda$-plane is chosen to the right-hand side of the rightmost singularity, i.e. $\chi> \lambda_0$ (see \cite{Reuter08} for more  details).
From Eq. (\ref{Eq8}) in  the thermodynamic limit $V\rightarrow\infty$ one finds the system pressure as  $p \equiv T \lambda_0$, since  $Q (T, \mu, V\rightarrow\infty) \sim \exp(\frac{p V}{T})$ by definition
 \cite{Huang}.  
Equation for $p$ in this limit becomes
 \begin{equation}\label{Eq9}
\hspace*{-1.5mm} p (T, \nu) = \pm T g \hspace*{-1.5mm} \int \hspace*{-1.5mm} \frac{d^3 k }{(2 \pi \hbar)^3}   \ln  \left[1 \pm \exp\left[ \frac{\mu - b p - e(k)}{T} \right] \right] ,
\end{equation}
where $e(k) \equiv \sqrt{k^2 + m^2}$ is the relativistic energy of a particle of mass $m$ and $\nu \equiv \mu - b p$ is  the shifted chemical potential.  Thus, Eq. (\ref{Eq9})  justifies the shift of the chemical potential $\mu$ suggested in Ref. \cite{Dirk91} to introduce the quantum hard-core repulsion in the grand canonical ensemble.  

In order to show that the pressure (\ref{Eq9}) is, indeed,  similar to the VdW EoS at low densities one has to find out the particle number density $n$,  to write the quantum  virial expansion and substitute in it the expression for density. Although this is a known subject, see, for instance, \cite{Huang} and a detailed analysis of quantum virial expansion made in Ref. \cite{KABugaev18}, for our purposes it is sufficient to consider the low density limit, i.e. if $\max \{\omega(k, p/T) \} \ll 1$. Then for the particle number density $n$ one finds  
\begin{eqnarray}\label{Eq10}
&& \hspace*{-9.9mm} n\equiv \frac{\partial p}{\partial \mu} = \frac{n_{id}(T, \nu)}{1+ b n_{id}(T, \nu)} ~\Rightarrow~ n_{id}(T, \nu) = \frac{n}{1- b n}\,, \\
 && \hspace*{-9.9mm}\, n_{id}(T, \nu) \equiv \frac{\partial p}{\partial \nu} =  g \hspace*{-1.5mm} \int \hspace*{-1.5mm}\frac{d^3 k  \, }{(2 \pi \hbar)^3} 
  \frac{\omega (k, p/T)}{1 \pm \omega (k, p/T)}  
\, \simeq ~ \nonumber\\
\label{Eq11}
&& \hspace*{-9.9mm} \simeq g \hspace*{-1.5mm}\int \hspace*{-1.5mm}\frac{d^3 k  \, \omega (k, p/T) }{(2 \pi \hbar)^3} \left[1 \mp \omega (k, p/T) + \omega (k, p/T)^2 ... \right],\,
\end{eqnarray}
where the particle number density of point-like particles $ n_{id}(T, \nu)$ is expanded in powers of the Boltzmann exponential $\omega (k, p/T)$. Similarly  expanding the $\ln(1\pm \omega(k, p/T))$ function in this limit, one obtains
\begin{eqnarray}\label{Eq12}
 && \hspace*{-9.9mm} p^{M1}(T, \nu) \equiv p_{id}(T, \nu)
  \simeq T g \hspace*{-1.5mm}\int \hspace*{-1.5mm}\frac{d^3 k  \, \omega (k, p/T) }{(2 \pi \hbar)^3}
   \times \nonumber \\
 && \hspace*{-9.9mm} \times  
  \left[1 \mp \frac{\omega (k, p/T)}{2} + \frac{\omega (k, p/T)^2}{3} ... \right] \simeq \\
  \label{Eq13}
  && \hspace*{-9.9mm} T n_{id}(T, \nu) \left[ 1 + a_2^{(0)}n_{id}(T, \nu)  + a_3^{(0)}n_{id}(T, \nu)^2 ...  \right] \,,
\end{eqnarray}
where the quantum virial  expansion of $p_{id}(T, \nu)$ for point-like particles (ideal gas) in terms of  quantum virial coefficients $a_j^{(0)}$ 
and powers of $n_{id}(T, \nu)$ is used.   The explicit formulas for $a_j^{(0)}$  can be found in \cite{Huang,KABugaev18}. 
However, for our purpose it is important that such an  expansion can be obtained from Eqs. (\ref{Eq9}) and  (\ref{Eq13}).  Substituting the right  Eq. (\ref{Eq10}) into expression for pressure  (\ref{Eq13}), one finally gets
\begin{eqnarray}\label{Eq14}
 && \hspace*{-11.1mm} p^{M1} 
  \simeq   \frac{T \,n}{1- b\, n} \left[ 1 +  \frac{a_2^{(0)} n}{1- b\, n}  + \frac{ a_3^{(0)}  n^2}{\left[1- b\, n\right]^2} \hspace*{0.01mm}+...  \right] ,\,
\end{eqnarray}
where the first term on the right-hand side of Eq. (\ref{Eq14}) is the Boltzmann gas pressure, while {  the other terms} are the quantum corrections. 
Since the coefficients  $a_j^{(0)}$ do not depend on $\nu$, but only on the temperature $T$  \cite{Huang,KABugaev18}, Eq. (\ref{Eq14}) is the quantum VdW EoS   in the canonical ensemble variables $T$ and $n$. 
%
%
\section{3. Going beyond the VdW approximation}\label{Sect3}
Above we have seen that in contrast to the claims of Ref. \cite{Typel16} there is no ambiguity in deriving the quantum gas pressure with the hard-core repulsion. Nevertheless, now
%
we modify the derivation of  the preceding  section in order to demonstrate how dangerous the outcome of the suggestion of  Ref. \cite{Typel16} to modify  the system volume via the effective degrees of freedom  may be.
For pedagogical  reasons, we consider only  an inclusion of the third classical virial coefficient into the quantum partition. Here we follow the suggestion of Ref. \cite{Typel16} to modify   the number of degrees of freedom, or  degeneracy factor $g$. Hence instead of Eq. (\ref{Eq5}) we write
\begin{eqnarray}\label{Eq15}
&&\hspace*{-7.7mm}
 \sum\limits_k  \ln  \left[1 \pm \omega (k, \lambda) \right] \rightarrow \tilde V G(\lambda) \hspace*{-1.4mm} \int \hspace*{-1.4mm}\frac{d^3 k  \, g}{(2 \pi \hbar)^3}   \ln  \left[1 \pm \omega (k, \lambda) \right] 
 \hspace*{-0.44mm} ,\, \,
\end{eqnarray}
where the additional  factor $G(\lambda)$ is introduced. Note that one could, of course,   introduce some complicated function $\tilde n (T, \mu)$ by  hand  and consider the dependence $G(\tilde n (T, \mu)) $ instead of $G(\lambda)$. At the end of  the derivation one could of course choose the former function $G$ in such a way that 
$\tilde n (T, \mu)$ would coincide with the particle number density $n$ of the system. 
Such a trick is employed in Ref. \cite{Vovch17} to generalize the Carnahan-Starling EoS  to the quantum case.  One should, however, remember that at this stage of the derivation neither the pressure of the system $p$ nor its derivatives (including $n$) are defined yet. Hence, if the factor $G(\lambda)$ is introduced into Eq. (\ref{Eq15}), then in the absence of additional assumptions, i.e. in  the simplest case, a mathematical consistency requires that $G(\lambda)$
{\it must depend on $\lambda$ only}.  
The practical reason for such a choice is that for dilute systems that are analysed in this section  the quantity $p/T$ is the particle number density of point-like particles (see Eq. (\ref{Eq13}) for M1 and below for this model). 

Assuming that the factor $G(\lambda)$ does not generate the rightmost singularity of the isobaric partition ${\cal Q} (T, \mu, \lambda)$, one can repeat the steps of deriving Eqs. (\ref{Eq6})-(\ref{Eq8}) for the hypothesis (\ref{Eq15}) and obtain the following result for  the system pressure in the thermodynamic limit
 \begin{equation}\label{Eq16}
\hspace*{-2.2mm} \frac{p (T, \nu)}{T} = \pm   G \hspace*{-1.1mm}\left(\frac{p}{T}\right) g \hspace*{-1.5mm}  \int \hspace*{-1.5mm} \frac{d^3 k }{(2 \pi \hbar)^3}   \ln\hspace*{-0.55mm}\left[1 \pm \exp\hspace*{-0.55mm}\left[ \frac{\nu - e(k)}{T} \right] \hspace*{-0.55mm}\right],
\end{equation}
where at the end of the calculations we used the explicit form of the function $\omega(k,p/T)$ and the same notation for the shifted  chemical potential $\nu\equiv \mu - bp$, as above.  Comparing Eqs.  (\ref{Eq9})   and  (\ref{Eq16}) for the system pressure, one concludes that 
the degeneracy factor is, indeed, modified  by the function   $G \left(\frac{p}{T}\right) $. It is convenient to represent Eq. (\ref{Eq16}) for an auxiliary density $\rho \equiv p/T$ in a compact form 
 \begin{equation}\label{Eq17}
\hspace*{-2.2mm} \rho 
 = \frac{ G \left(\rho\right) }{T} p_{id} (T, \nu) , ~\Rightarrow ~ n_{id}(T, \nu) \equiv \frac{\partial p_{id}(T, \nu)}{\partial \nu}\,,
\end{equation}
where the pressure of point-like particles $p_{id} (T, \nu)$ with the shifted  chemical potential $\nu$ is given by the right-hand side of Eq.  
(\ref{Eq9}), since the pressure of M1 
is indeed the  pressure of an ideal gas with the shifted value for the chemical potential $\nu$.
Hence, the definition of the particle number density of such particles $n_{id}(T, \nu)$ coincides with the  right-hand side of Eq. (\ref{Eq11}). 
Then the particle number density of the present model (M2 hereafter) is 
\begin{eqnarray}\label{Eq18}
&& \hspace*{-9.9mm} n\equiv \frac{\partial p}{\partial \mu} = \frac{n_{id}(T, \nu)}{G(\rho)^{-1}+ b n_{id}(T, \nu) - \frac{G^\prime(\rho) }{G(\rho) } \frac{p_{id}(T, \nu)}{ T}} \, ,
\end{eqnarray}
where {the notation $G^\prime(\rho) = \frac{d G(\rho)}{d \rho}$ is introduced.}

For dilute systems one can substitute the virial expansion (\ref{Eq13}) for $p_{id}(T, \nu)$ into Eq.  (\ref{Eq18}) and express the density of point-like particles as
\begin{eqnarray}\label{Eq19}
&& \hspace*{-9.9mm} n_{id}\simeq \frac{G(\rho)^{-1} n}{1 - b n + \frac{G^\prime(\rho) }{G(\rho) }n [1+ a_2^{(0)}n_{id}  + a_3^{(0)}n_{id}^2 +... ]} \, .
\end{eqnarray}
Expansion of  $p_{id}(T, \nu)$ in Eq. (\ref{Eq17})  according to  Eq. (\ref{Eq13})
with the simultaneous substitution  of  (\ref{Eq19})   into (\ref{Eq17}) allows one to write the pressure of M2 as
\begin{eqnarray}\label{Eq20}
&& \hspace*{-9.9mm} p^{M2} \simeq \frac{ n T[1+ a_2^{(0)}n_{id}  + a_3^{(0)}n_{id}^2 +... ]}{1 - b n + \frac{G^\prime(\rho) }{G(\rho) }n [1+ a_2^{(0)}n_{id}  + a_3^{(0)}n_{id}^2 +... ]} \, .
\end{eqnarray}
In order to  reveal  the effect of  modifying the  degeneracy factor on the quantum virial expansion of M2  let us consider one of the simplest choices   for  the function $G(\rho)$ 
\begin{eqnarray}\label{Eq21}
&& \hspace*{-7.7mm}  \frac{d G}{d \rho} = - \frac{c_3 \rho}{1+c_4 \rho} ~\hspace*{-0.55mm}\Rightarrow~ \hspace*{-0.55mm}G  = \exp\hspace*{-0.55mm}\left[  \frac{c_3}{c_4^2}\ln |1+c_4 \rho | - \frac{c_3}{c_4}\rho\right], ~\,
\end{eqnarray}
which contains two constants $c_3$ and $c_4$ whose meaning will be clear in a moment. This example allows us to get the quantum third virial coefficient for a  dilute system, by writing $G^\prime (\rho) \simeq -c_3 \rho \simeq -c_3 n~\Rightarrow ~ G(\rho)^{-1} \simeq 1 + \frac{c_3}{2} \rho^2 \simeq 1$, where in the last step we approximated the M2 pressure by the ideal gas one $nT$. Furthermore, expanding the denominator 
of Eq. (\ref{Eq20}) with respect to  the powers of $G^\prime (\rho)/G(\rho)$, one gets
\begin{eqnarray}\label{Eq22}
 && \hspace*{-11.1mm} 
 \frac{1}{1- b\, n} \left\{ 1 +  \frac{c_3 n^2[1+...]}{1- b\, n}  + \frac{ c_3^2  n^4 [1+...]^2}{\left[1- b\, n\right]^2} \hspace*{0.01mm}+...  \right\} ,\,
\end{eqnarray}
where the shorthand notation $[1+...] \equiv [1+ a_2^{(0)}n_{id}  + a_3^{(0)}n_{id}^2 +... ] \simeq 1$ is introduced. With the help of this result and Eq. (\ref{Eq14})  one can establish the following relation for the M2 pressure (\ref{Eq20})
\begin{eqnarray}\label{Eq23}
&& \hspace*{-11.mm} p^{M2} \hspace*{-0.55mm}\simeq p^{M1} \hspace*{-0.55mm}\left\{ 1 +  \frac{c_3 n^2[1+...]}{1- b\, n}  + \frac{ c_3^2  n^4 [1+...]^2}{\left[1- b\, n\right]^2} \hspace*{-0.55mm}+...  \right\} ,\, 
\end{eqnarray}
which allows us to easily find the relation between the  quantum third virial coefficients of M1 and M2. Indeed, since the virial expansion for any pressure is defined as $p = nT[1+a_2n+a_3n^2 + ...]$, from Eq. (\ref{Eq23}) one concludes that the term $c_3n^2$ in the brackets of (\ref{Eq23})  cannot  modify the second virial coefficient $a_2^{M1}$ of M1, but will shift $a_3^{M1}$ and all higher order coefficients.
Hence, one can write
\begin{eqnarray}\label{Eq24}
&& \hspace*{-11.mm} a_3^{M2} = a_3^{M1} + c_3  =  a_3^{(0)} + 2a_2^{(0)} b + b^2 + c_3,\, 
\end{eqnarray}
where for $ a_3^{M1} $ the result of Ref. \cite{KABugaev18} is used. It is well known that the classical VdW EoS provides the wrong value for the  third virial coefficient of hard spheres $b^2 = 16 V_0^2$ instead of $10 V_0^2$ \cite{Huang}. Hence, one can use the parameter $c_3$ to cure this problem. If one chooses $c_3 = - 6V_0^2 = -\frac{3}{8}b^2$, then  the M2 third virial coefficient  (\ref{Eq24}) will have the correct classical limit, when the quantum corrections proportional to $a_3^{(0)}$ and $a_2^{(0)}$ are negligible, i.e. at high values of $T$ 
\cite{Huang,KABugaev18}. Such a choice, however, would lead to a huge enhancement 
of the degeneracy factor, if  in Eq.  (\ref{Eq21}) the coefficient $c_4=0$ is  negligible.

From the right  Eq. (\ref{Eq21}) one can see that  for  a positive value of the coefficient $c_4$  the resulting EoS has a structure that is similar to the hard-core repulsion for  large values of $\rho$, i.e. $G(\rho) \sim \exp\left[ -\frac{c_3}{c_4} \rho\right]$, but for $c_3 = - 6V_0^2 < 0 $ and $c_4>0$ instead of  a repulsion one has an attraction. Hence, in the limit of Boltzmann statistics (high temperatures) the coefficient $c_4$ can be fixed to reduce  the value of the excluded  volume of particles from $b=4V_0$  to their  eigen volume $V_0$. In other words, writing $b+\frac{c_3}{c_4}=  4V_0- \frac{6V_0^2}{c_4} = V_0$, one finds $c_4=2 V_0 = 0.5 b$. 

At first glance such a model looks better than the VdW EoS, but, in fact, it leads to the limiting temperature $T_{lim} (\nu)$ as a function of the shifted chemical potential $\nu$. The reason for such a behaviour   can be seen from  the left Eq. (\ref{Eq17}). It is evident that for  a large value of $p_{id}(T, \nu)/T$ the right-hand side of this equation  at $\rho=0$ can be so large, i.e. $G(0)p_{id}(T, \nu)/T \gg 1$,  that the monotonically growing function $G(\rho)p_{id}(T, \nu)/T$ of $\rho$ would  never cross or  be tangential to the straight line $\rho$ and hence there would be no solution of the left  Eq. (\ref{Eq17}). The critical value of the variable $p_{id}(T, \nu)/T$ for which such a solution exists can be found from the  condition that the functions $G(\rho)p_{id}(T, \nu)/T$ and $\rho$ are tangential to each other at the point $\rho=\rho_t \equiv - [2 c_3]^{-1}[c_4 + \sqrt{c_4^2 - 4c_3}] $, which for  the coefficients   $c_3 = - 6V_0^2$ and $c_4 = 2 V_0$ discussed above is $\rho_t \simeq 0.6067 V_0^{-1}$.  Solving  the left  Eq. (\ref{Eq17}) for $\rho=\rho_t$, one obtains the following inequality  for the M2 parameters
\begin{eqnarray}\label{Eq25}
 && \hspace*{-0.11mm} p_{id}(T, \nu) \le 0.3237\, {T}\,{V_0}^{-1}\,
 .\,
\end{eqnarray}
Since the ideal gas pressure $p_{id}(T, \nu)$ is a monotonously increasing function of the variables $T$ and $\nu$, then the solution of inequality (\ref{Eq25}) is $T \le T_{lim} (\nu)$, where it is assumed that for $T=T_{lim} (\nu)$  the inequality (\ref{Eq25}) is obeyed  as equality.  Apparently, the same conclusion is valid for a zero value of the coefficient $c_4$, since for $c_3= -6 V_0^2 <0$ there is a strong enhancement of pressure due to $G(\rho)$.  

Note that the existence of limiting temperature for  the quantum VdW EoS of pion gas with both the hard-core repulsion and attraction was recently  found in Ref. \cite{Pober15}.  However, from the example considered above one sees that such an extravagant  behaviour  is common for both types of quantum statistics and it is generated by the enhancement of the system degeneracy factor. 

For negative values of  the coefficients $c_4$ and $c_3= -6 V_0^2$ the limiting temperature does not exist, since in this case one, indeed, has  a suppression of the degeneracy factor. However, such a model is also defective, since in the limit of Boltzmann statistics  for high values of $\rho$ in addition to the usual hard-core repulsion due to excluded volume $b$, the $G(\rho)$ factor provides an { extra}  suppression, which { increases  the} resulting excluded  volume $b+\frac{c_3}{c_4} >  4V_0$. { This  means that  such  EoS  becomes invalid at lower packing fractions  than the classical VdW one.}

The example considered above clearly demonstrates the fact that improvement of the hard-core repulsion interaction by modifying the degeneracy factor leads to severe problems of the obtained EoS.  Evidently, these problems are generated by the  trick  to move the free energy of interacting particles  from the statistical operator to  the factor modifying the degeneracy of particles.  Therefore, neither the statement of non-uniqueness of the procedure to include the  hard-core repulsion into quantum gases \cite{Typel16} nor the generalization of the Carnahan-Starling EoS to quantum systems  that  uses  a trick  to modify the degeneracy factor \cite{Vovch17}  look  at the moment consistent with the derivation based on evaluating the quantum partition. 
%
%
\section{4. Excluded volume of multicomponent systems}\label{Sect4}
In order to derive the quantum EoS for several sorts of particles with different hard-core radii we employ an entirely new approach. First, we evaluate the excluded volume of  the multicomponent system within a self-consistent treatment for 3-dimensional hard spheres. Second, using the freedom that
the excluded volume does not fix the 3-rd, 4-th and higher order virial coefficients, we introduce the phenomenological parameters that allow us
to account for the higher order virial coefficients not only for  hard spheres, but for any convex particles. Finally, we generalize this scheme to any dimension $D\ge2$.

Consider now $K \ge 1$ sorts of quantum particles of different  statistics with the masses $m_K$, chemical potentials $\mu_K$
and hard-core radii $R_K$. Assume that the total particle number of $K$-th sort  is $N_K=n_{K,1}+n_{K,2}+n_{K,3}+...$
and they have energy $E_K=e_{K,1}n_{K,1}+e_{K,2}n_{K,2}+e_{K,3}n_{K,3}+...$, while the total number of particles is $N_{tot}=\sum\limits_K N_K$. Here $n_{K,l}$ is the number of particles of sort $K$ occupying the energy level  $e_{K,l}$. Assuming that the hard-core repulsion can be taken into account via the excluded volume $V_{excl}$,
one can write a multicomponent analog of partition (\ref{Eq1}) as

\begin{eqnarray}\label{Eq26}
&&\hspace*{-8.8mm}Q (T, \{\mu_K\}, V) = \nonumber \\
&&\hspace*{-8.8mm}\sum\limits_K\sum\limits_{N_K=0}^\infty \sum\limits_{\sum\limits_{l=1}[ \, n_{K,l}]=N_K} \hspace*{-4.4mm}e^{\left[ \frac{\mu_K N_K - E_K}{T} \right]} \Theta\left(V - V_{excl}  \right),\, 
\end{eqnarray}
{ here the  excluded volume of all pairs  taken per particle is} 
\begin{eqnarray}\label{Eq27}
&&\hspace*{-7.7mm}
{V}_{excl}=  \frac{1}{N_{tot}}{\sum_{K,L} N_K \frac{2}{3} \pi (R_K+R_L)^3 N_L} \,.
\end{eqnarray}
Since an exact  evaluation of the partition (\ref{Eq26}) with excluded volume (\ref{Eq27}) is very complicated, we are going to simplify Eq.(\ref{Eq27}) by introducing statistical average quantities. Writing explicitly the binomial in Eq. (\ref{Eq27}) and regrouping the terms with powers of $R_K$ one finds
\begin{eqnarray}\label{Eq27n}
&&\hspace*{-11.mm}
{V}_{excl}=   \sum_{K}N_K\left[ v_K +\frac{s_K}{2}  \cdot \overline{R} +\frac{c_K}{2}   \cdot\overline{R^2} \right] = \\
&&\hspace*{-11.mm}
\label{Eq28}
 =\sum_{K}N_K\left[ v^*_K +\frac{s_K}{2}  \cdot \overline{R} +\frac{c_K}{2}    \cdot\overline{R^2} +\frac{4}{3}\pi u_K\cdot\overline{R^3}  \right] ,\,
\end{eqnarray}
where in the first equality above  we introduced the eigen volume $v_K =\frac{4}{3}\pi R_K^3 $, eigen surface $s_K ={4}\pi R_K^2 $ and eigen (double) perimeter $c_K ={4}\pi R_K $ of $K$-th sort of  spheres  and the  powers of hard-core radii $\overline{R^n} \equiv N_{tot}^{-1} \sum_{L} R^n_L N_L$.
In the second equality above we introduced the closest packing volume of each sort of sphere $v^*_K \equiv v_K/q$ with $q = \frac{\pi}{3\sqrt{2}}\simeq 0.741$ and the compensative ``mean-field" term which is proportional to  $u_K = 1-1/q \simeq - 0.35$.
With these notations 
Eq. (\ref{Eq28}) is exact, but now we introduce the self-consistent approximation for it by replacing the exact averaging in the excluded volume expression (\ref{Eq28}) with the statistical ones (for  $n=1, 2, 3$)
\begin{eqnarray}\label{Eq29}
&&\hspace*{-11.mm}
	\overline{R^n} \equiv N_{tot}^{-1} \sum_{L} R^n_L N_L  \rightarrow \frac{\sum_{L}R^n_L \left\langle N_L\right\rangle}{\sum_{L}\left\langle N_L\right\rangle} ,\,
\end{eqnarray}
where $\langle N_L \rangle \equiv T \frac{\partial}{\partial \mu_L} \ln Q (T, \lbrace \mu_k \rbrace, V)$ is the mean number of particles of $L$-th sort found from the partition (\ref{Eq26}). Assuming that the partition (\ref{Eq28}) under approximation (\ref{Eq29}) is known, then the quantity $\overline{R^n} $ can be found from it as 
\begin{eqnarray}\label{Eq30}
&&\hspace*{-11.mm}
	\overline{R^n}  = \frac{ \sum_{L = 1} R^n_L\frac{\partial}{\partial \mu_L}\ln Q(T,\left\lbrace \mu_k \right\rbrace, V)}{\sum_{L =1} \frac{\partial}{\partial \mu_L}\ln Q (T, \lbrace \mu_k \rbrace,V)} .\,
\end{eqnarray}
Under approximation (\ref{Eq29}) the { excluded volume  (\ref{Eq28})} becomes a linear function of $N_K$ and, therefore, 
the  partition (\ref{Eq26})  can be found  by repeating  the formal steps of deriving  Eqs. (\ref{Eq2})-(\ref{Eq8}). Moreover, 
in  the thermodynamic limit 
the formal expression for  the approximated partition (\ref{Eq26}) coincides with Eq. (\ref{Eq8}), i.e. $Q (T, \{\mu_K\}, V\rightarrow \infty) \sim \exp{\left[\frac{p V}{T}\right]}$, where  the pressure is 
\vspace*{-3mm}
 \begin{eqnarray}\label{Eq31}
\hspace*{-1.4mm} p ^{M3}=  T \sum\limits_K g_K \hspace*{-1.4mm} \int \hspace*{-1.4mm} \frac{a_K d^3 k }{(2 \pi \hbar)^3}   \ln \hspace*{-1.1mm}  \left[1 +  \frac{\exp\left[ \frac{\nu_K^0 - e_K(k)}{T}  \right] }{ a_K}\right] ,\\
\label{Eq32}
\hspace*{-1.4mm} \nu_K^0 \equiv \mu_K - \left[v^*_K +\frac{ s_K}{2} \overline{R} +\frac{c_K}{2} \overline{R^2} +\frac{4}{3}\pi u_K\overline{R^3} \right]p^{M3} .~
\end{eqnarray}
\vspace*{1mm}
In Eq. (\ref{Eq31}) the shifted chemical potential (\ref{Eq32}) is set for each sort of particle, $e_K(k)=\sqrt{k^2 + m^2_K}$ denotes the energy of $K$-th sort of particle  and the parameter $a_K =-1$ should be taken for the BE statistics, while for the FD one { it  must be set to} $a_K = 1$. Calculating the partial derivative $\frac{\partial p^{M3}}{\partial \mu_K}$ from Eq. (\ref{Eq31}) as a derivative of the implicit function $p^{M3}$, one can write the quantity 
$\overline{R^n}$  (\ref{Eq30}) as 
\begin{eqnarray}\label{Eq33}
&& \hspace*{-11.mm}\overline{R^n} =  \frac{ \sum_K \,R_K^n   n_{id}^K (T, \nu_K^0)  }{ \sum_L  n_{id}^K (T, \nu_K^0)  )}, 
~ {\rm for}\, \,~ n_{id}^K (T, \nu_K^0)  \equiv \frac{\partial p^{M3}}{\partial \nu_K^0} ,\,\,  
\end{eqnarray}
where we introduced the density of point-like particles $n_{id}^K (T, \nu_K^0)$  of sort $K$  which is just Eq. (\ref{Eq11}) with the following replacements $g \rightarrow g_K$,  $m\rightarrow m_K$ and $\nu \rightarrow \nu_K^0$.
Eqs.  (\ref{Eq31})--(\ref{Eq33}) of model 3 (M3) form a closed system of equations for the quantum  VdW EoS of hard spheres. 
The M3 pressure (\ref{Eq31}) is a quantum generalization of the approach of Refs. \cite{Dillmann:91,LFK:94}, but in contrast to Refs. \cite{Dillmann:91,LFK:94} and their followers,  the coefficients of surface $0.5 \overline{R^1}$ and curvature  $0.5 \overline{R^2}$ tensions
are not the fitting parameters, but are defined by  the system (\ref{Eq31})--(\ref{Eq33}). 

Our next step is to generalize the system (\ref{Eq31})--(\ref{Eq33}) to convex particles of arbitrary shape and to extend it  to higher densities.  For these aims one has to  not only consider the eigen volume of convex particle of  $K$-th sort $v^c_K$, their eigen surface $s_K^c$, two mean curvature radii $\bar{r}_{K,1}$ and $\bar{r}_{K,2} \ge \bar{r}_{K,1}$, their  corresponding (doubled) perimeters  $c_{K,1}^c=4 \pi \bar{r}_{K,1}$ and $c_{K,2}^c=4 \pi \bar{r}_{K,2}$ (for more details see Ref. \cite{Isihara}), but also to modify  their weights in the expressions  for the excluded volume (\ref{Eq28}) and for the chemical potential  $\nu_K^0$ in (\ref{Eq32}). To demonstrate this let us consider the famous Isihara-Hadwiger (IH) formula \cite{Isihara1,Hadw1,Isihara} for the classical second virial coefficient of two identical convex particles 
\begin{equation}\label{Eq34n}
2V_{excl}^c \equiv 2v^c_1+ s_1^c(\bar{r}_{1,1}+\bar{r}_{1,2}) ,
\end{equation}
where we used the first subscript 1 for both particles. Introducing now the equivalent sphere radius as $s_1 \equiv 4 \pi R_s^2= s(R_s)$, one can identically rewrite the IH formula with the help of two auxiliary constants $A, B\in [0; 1]$ as 
\begin{eqnarray}\label{Eq35n}
2V_{excl}^c & \equiv & \tilde v_1 + \tilde \alpha_2 s(\bar{r}_{1,1})R_s + (1-A)c_{1,1}^c R_s^2 \nonumber \\
&+ &\tilde v_2
+ \tilde \alpha_2 s(\bar{r}_{1,2})R_s + (1-A) c_{1,2}^c R_s^2 ,
\end{eqnarray}
where the notations
\begin{eqnarray}\label{Eq36n}
\tilde v_1 & = & v^c_1 -  B s(\bar{r}_{1,1}) R_s (\tilde \alpha_1 - \tilde \alpha_2) , \\
\tilde v_2 & = & v^c_1 + (1+B) s(\bar{r}_{1,1}) R_s (\tilde \alpha_1 - \tilde \alpha_2) ,\\
\label{Eq37n}
\tilde \alpha_k & \equiv & A\frac{R_s}{\bar{r}_{1,k}} \quad {\rm with} \quad  k=1, 2 ,
\end{eqnarray}
are used.  This example shows us that the IH formula has a structure of the excluded volume of   Eq.  (\ref{Eq27n}) for $K=1, 2$ and $N_K=1$, but with different numerical weights, if one identifies $R_s^n$ with $\overline{R^n}$.  Also this  example demonstrates us the meaning 
of the quantities entering Eq. (\ref{Eq35n}): thus, the eigen surfaces $s(\bar{r}_{1,k}) \Leftrightarrow s_k$ and eigen  (double) perimeters  $c_{1,k} \Leftrightarrow c_k $ are related 
to the mean curvature radii, while the effective volumes $\tilde v_K$ can, in general, differ from the eigen volume $v_1^c$.
Moreover,   it is clear that 
using  the  quantities  $v_K^*$, $s_K$, $c_K$,  $u_k$  and  their  weights in Eq. (\ref{Eq32}) as the adjusting parameters,
one has sufficient number of  parameters  to reproduce $V_{excl}$ of the mixture of convex particles of different shapes and sizes.

Although the system (\ref{Eq31})--(\ref{Eq33}) is derived under a self-consistent approximation, for the practical purposes it looks rather complicated and, hence, we simultaneously simplify it and extrapolate to higher densities.  First, we recall that for low densities one can approximate the pressure (\ref{Eq31}) using the lowest density term in the virial expansion as 
\begin{equation}\label{Eq39n}
p^{M3} \equiv \sum_K  p_{id}^K (T, \nu_K^0) \simeq T \sum_K n_{id}^K (T, \nu_K^0) ,
\end{equation}
where we introduced the partial pressure for each sort of  particle $p_{id}^K (T, \nu_K^0) $ and approximated it by the ideal gas one. For such an approximation the terms $\overline{R^n} p^{M3}$ appearing in the expression for the shifted chemical potential (\ref{Eq32}) can be approximated as 
\begin{equation}\label{Eq40n}
\hspace*{-1.1mm}\overline{R^n} p^{M3} \simeq T\sum_K \,R_K^n   n_{id}^K (T, \nu_K^0)  \simeq  \sum_K \,R_K^n   p_{id}^K (T, \nu_K^0) ,
\end{equation}
where in the second step of the derivation we reverted the 
approximation  
$T  n_{id}^K (T, \nu_K^0) \simeq p_{id}^K (T, \nu_K^0)$ used before.
Apparently, such an approximation does not change the second virial coefficient of  $p^{M3}$, since the latter ones are defined by  the terms  $\overline{R^n} p^{M3}$ and, hence, their approximation 
modifies  the higher order virial coefficients, { which now will be  modified  by redefining the chemical potentials $\nu_K^n$ in the terms 
$\overline{R^n} p^{M3} \simeq \sum_K \,R_K^n   p_{id}^K (T, \nu_K^n)$. {\it Since for the Boltzmann statistics, i.e. for  $a_K \rightarrow \infty$, the last result follows directly from Eqs. (\ref{Eq33}) (see also Ref. \cite{ISCT2019} for a rigorous derivation), we adopt it and extrapolate it to all densities in the quantum case.}
Taking into account  
all the modifications discussed  above  the model four  (M4) pressure is}  
 \begin{eqnarray}\label{Eq34}
&&\hspace*{-7.4mm}p ^{M4}=   \sum\limits_K  p_{id}^K (T, \nu_K^0)  ,\\
\label{Eq35}
&&\hspace*{-7.4mm} \nu_K^0 \equiv \mu_K - \left[v^*_Kp^{M4} +{s_K}f_1 + {c_K}f_2 + u_K f_3 \right] ,\\
\label{Eq36}
&&\hspace*{-7.4mm} f_n \equiv  A_n \sum\limits_K  R_K^n \,\,p_{id}^K (T, \nu_K^n), ~{\rm for}~ n=1, 2, 3,\,\\
\label{Eq37}
&&\hspace*{-7.7mm} \nu_K^1 \equiv \nu_K^0 - \Delta_K^1 {s_K}f_1,\, ~\nu_K^2 \equiv \nu_K^1 - \Delta_K^2  {c_K}f_2,\,\\
\label{Eq38}
&&\hspace*{-7.7mm} \nu_K^3 \equiv \nu_K^2 - \Delta_K^3  {u_K}f_3,\,
\end{eqnarray}
{where the constants $A_n$ (with $A_{n\neq3} > 0$) and $\Delta_K^n > 0$ are,  respectively, the common and individual (for $K$-th sort of particles) weights with which the average quantity $\overline{R^n} p^{M3}$}
 enters  the expressions of chemical potentials $\nu_K^n$.  
 Note that in nuclear physics the leptodermous expansion of the binding energy of large nuclei   contains similar terms  \cite{Sagun19} and, moreover, the mean-field one $u_K f_3$ is often  called  the Gaussian curvature \cite{Sagun19}. 
 
 The positive constants $\Delta_K^n > 0$ are introduced in order to account for the third, fourth and higher order virial coefficients of classical hard spheres.  Such an approach leads to the hierarchy of  partial pressures of point-like particles 
\begin{equation}\label{Eq47n}
p_{id}^K (T, \nu_K^{n-1}) > p_{id}^K (T, \nu_K^n)  \quad {\rm for} \quad n = 1, 2,
\end{equation}
since for a positive  value of coefficient  $A_n$ one finds $ \nu_K^{n-1} >  \nu_K^n$.  It is evident that the higher  the pressure is, the higher  the density is  and the stronger the inequalities   (\ref{Eq47n})
 and $ \nu_K^{n-1} >  \nu_K^n$  are for $n = 1, 2$.  Therefore, at high particle number  densities one should expect that for properly chosen coefficients $A_n$ and $\Delta^K_n$, and  for positive values of  $v_K^*$, $s_K$ and  $c_K $ the following set of inequalities  
\begin{equation}\label{Eq48n}
v_K^* p^{M4} \gg s_K f_1 \gg c_K f_2 ,
\end{equation}
should appear. Moreover, {for any sign of $A_3$ coefficient,  the inequality $ \nu_K^{0} \gg  \nu_K^3$ can be easily established at high densities  due to a strong suppression of  $p_{id}^K (T, \nu_K^{n})$, where $n=1, 2$ compared to $p_{id}^K (T, \nu_K^0)$, and, hence,  in this limit the M4 EoS will reproduce the dense packing behaviour.} 
 
Therefore, choosing the coefficients $A_n$ and $\Delta_n^K$ from fitting the pressure of classical hard spheres in the whole gaseous phase, one can substitute them into the M4 equations (\ref{Eq34})-(\ref{Eq38}) and  get the quantum EoS for multicomponent mixture of hard convex particles  for a dimension $D=3$.  Our confidence is based on the fact that for the one-component case  with a truncated system (\ref{Eq34})-(\ref{Eq38}), which accounts only  for the pressure $p^{M4}$ and for $f_1$, one could choose $\Delta_1^1=0.245$ ($A_1=2$ was fixed, while $A_{n>1}=0$) to simultaneously reproduce the second,  third and fourth   virial coefficient of classical hard spheres \cite{KABugaev18,IST2018}. Such an EoS provides  a very good description of nuclear \cite{Aleksei18}, hadronic  \cite{IST2018} and neutron matter \cite{Sagun:NS18} EoS properties with the minimal number  of adjustable parameters. 
{Moreover, it can be shown numerically \cite{ISCT2019} that the standard  compressibility factor $Z \equiv p/ (n T)$ (here $p$ denotes pressure and $n$ is the particle number density) of one- and two-component mixtures of hard spheres and hard discs are excellently  reproduced by the system  (42)-(46)  for the whole gaseous phase up to the transition to the solid phase.
}

Apparently, the whole derivation of M3 and M4 pressures can be straightforwardly extended to arbitrary dimension $D\ge 2$ by replacing the  integration measure  
\begin{equation}
\frac{d^3 k}{(2 \pi \hbar)^3} \rightarrow \frac{d^D k}{(2 \pi \hbar)^D}
\end{equation}
in the corresponding expressions  and by replacing  the $3$-dimensional eigen volume $v_K$ by the $D$-dimensional eigen hyper-volume  $v_K^D$, the $2$-dimensional eigen surface $s_K$ by the $(D-1)$-dimensional eigen 
hyper-surface and so on in them,  to have at the end $D+1$ equations similar to the system (\ref{Eq34})-(\ref{Eq38}). Since the virial coefficients of classical  hard spheres are known for  higher dimensions \cite{VirialC}, they can be used in the system (\ref{Eq34})-(\ref{Eq38}).
Having  the virial coefficients for the $D$-dimensional classical hard convex particles  one can determine their parameters $v_K^*$, $s_K$,   $c_K$ etc from the fitting pressure $p$ or the compressibility factor and write the all necessary parameters for  the system (\ref{Eq34})-(\ref{Eq38}).

\section{Conclusions} \label{Conclusions}
In this work we discussed the necessary conditions to derive the quantum VdW EoS with hard-core repulsion directly from the quantum partition.  Using  a plausible example it is shown that an alternative way to account for the hard-core repulsion suggested in Ref. \cite{Typel16} leads to severe  inconsistencies. The multicomponent formulation of the quantum VdW EoS with hard-core repulsion is derived within the self-consisting approximation. For practical applications  it is simplified,  extrapolated to higher densities and also generalized to the case of  hard convex bodies. {\it For the first time, the suggested approach allows one to treat the hard spheres and hard convex bodies of various  dimensions $D\ge 2$ on the same footing.}
The model parameters can be determined from the best description of classical systems and then used in the quantum EoS.  Since this approach is 
formulated solely in terms of the grand canonical variables, it can be used to model the properties of many physical systems with non-conserved number of particles.  In particular,  it can be used for the mixtures of constituents with very  different physical properties like a mixture of hadrons, nuclei and bags of quark-gluon plasma which are hard to model within the other approaches. \\

\noindent
{\bf Acknowledgments.}
The author appreciates the valuable comments of  I.P. Yakimenko, V. Vovchenko, L. Bravina, B. Grinyuk,  E. Zabrodin and N. Yakovenko and the  partial support by the National Academy of Sciences of Ukraine (project No. 0118U003197). Also the author  is grateful to the COST Action CA15213 ``THOR" for supporting his networking.


\begin{thebibliography}{99}

\bibitem{VDWeos}
%
 J. D. van der Waals,  Z. Phys. Chem. {\bf 5}, 133  (1889).

\bibitem{Ref2}
 J. P. Hansen and I. R. McDonald, Theory of Simple Fluids (Academic Press, Amsterdam, 2006).

\bibitem{Ref3}
 Theory and Simulation of Hard Sphere Fluids and Related Systems, Lect. Notes Phys. Vol. 753, edited by A. Mulero 
 (Springer-Verlag, Berlin, 2008).
 

\bibitem{Sagun14}
%
 V. V. Sagun, A. I. Ivanytskyi, K. A. Bugaev and I. N. Mishustin,
 Nucl. Phys. A {\bf 924}, 24  (2014) and references therein. 
 
 \bibitem{Aleksei18}
%
 A.~I.~Ivanytskyi, K.~A.~Bugaev, V.~V.~Sagun, L.~V.~Bravina and E.~E.~Zabrodin,
 Phys. Rev.  C {\bf 97},  064905 (2018). 
  
 \bibitem{IST2018}
%
 V. V. Sagun et al., 
 Eur. Phys. J. A {\bf 54},   100  (2018). 

 \bibitem{IST2018b}
 K. A. Bugaev et al., 
Nucl. Phys. A {\bf 970}, 133 (2018).


 \bibitem{Universe2019}
%
K.~A.~Bugaev et al., 
Universe {\bf 5},  00063, 1 (2019) and references therein.
 
\bibitem{3CEP}
%
  K. A. Bugaev et al., 
 Phys. Part. Nucl. Lett. {\bf 15}, 210 (2018).

\bibitem{Vovch17}
%
 V. Vovchenko,
 Phys. Rev. C 96, 015206 (2017).

\bibitem{CSeos}
%
 N. F. Carnahan, K.E. Starling, 
 J. Chem. Phys. 51, 635 (1969).



\bibitem{KABugaev18}
%
 K. A. Bugaev, A. I. Ivanytskyi, V. V. Sagun, E. G. Nikonov and G. M. Zinovjev,
 Ukr. J. Phys.  63,  863 (2018).



\bibitem{Dirk91}
%
 {D. H.  Rischke,  M. I.  Gorenstein,  H.  St\"ocker and W.  Greiner}, 
 {Z. Phys. C} {\bf 51},  {485} (1991).


\bibitem{Typel16}
%
 S. Typel, Eur. Phys. J. A {\bf 52},   16 (2016).

\bibitem{Qvdw1}
%
  V. Vovchenko, D. V. Anchishkin, and M. I. Gorenstein, Phys. Rev. C 91, 064314 (2015).

\bibitem{Qvdw2}
%
 K. Redlich and K. Zalewski, Acta Phys. Polon. B 47, 1943 (2016).

\bibitem{RelVDW1}
%
 K. A. Bugaev, M. I. Gorenstein, H. St\"ocker and W. Greiner,
 Phys. Lett. B {\bf 485},  121  (2000).

\bibitem{RelVDW2}
%
see 
 K. A. Bugaev, 
 {Nucl. Phys. A} {\bf  807},    251 (2008).

\bibitem{Reuter08}
%
for an appropriate review see  
K. A. Bugaev and P. T. Reuter,
 {Ukr. J. Phys.} {\bf 52},  489  (2007) and references therein. 

\bibitem{Huang}
%
 K. Huang,  {\it Statistical Mechanics} (Wiley \& Sons, 1967).

\bibitem{Pober15}
%
 R. V. Poberezhnyuk, V. Vovchenko, D. V. Anchishkin and M. I. Gorenstein, 
 J. Phys. G {\bf 43},  095105 (2016).

\bibitem{Dillmann:91} 
%
 A. Dillmann and G. E. Meier, 
 J. Chem. Phys. {\bf 94}, 3872 (1991). 

\bibitem{LFK:94} 
%
 A. Laaksonen, I. J. Ford, and M. Kulmala,
 Phys. Rev. E {\bf 49}, 5517 (1994).

\bibitem{Isihara1}
	A. Isihara, 
	J. Chem. Phys. {\bf 18},  1446 (1950).
	
\bibitem{Hadw1}
	H. Hadwiger,
	Mh. Math. {\bf  54}, 345 (1950).


\bibitem{Isihara}
%
 A. Isihara, {\it Statistical physics} (Academic Press, New York, 1971).



\bibitem{ISCT2019}
%
N. S. Yakovenko, K. A. Bugaev, L. V.  Bravina  and  E. E.  Zabrodin,
{\it The concept of induced surface and curvature tensions and a unified description of the gas of hard discs and hard spheres} (in preparation).

\bibitem{Sagun19}
%
V. V. Sagun, K. A. Bugaev  and  A. I. Ivanytskyi, 
 arXiv:1904.05955v1 [nucl-th].  

\bibitem{Sagun:NS18} 
   V.~V.~Sagun, I.~Lopes and A.~I.~Ivanytskyi,
   Astrophys.\ J.\  {\bf 871}, no. 2, 157 (2019).


\bibitem{VirialC}
%
 J. G. Loeser, Zh. Zhen, S. Kais and D. R. Herschbach,
 J. Chem. Phys. {\bf 95},  4525 (1991).



\end{thebibliography}
\end{document}